\documentclass[
    ,final            
  ]
  {aipproc}

\layoutstyle{6x9}

\begin{document}

\title{The {\em AstraLux} large M dwarf survey\footnote{Based on observations collected at the Centro Astron\'omico Hispano Alem\'an (CAHA) at Calar Alto, operated jointly by the Max-Planck Institut f\"ur Astronomie and the Instituto de Astrof\'isica de Andaluc\'ia (CSIC).}}


\author{Felix Hormuth}{
  address={Max-Planck-Institute for Astronomy, K\"onigstuhl 17, 69117 Heidelberg, Germany},
  altaddress={Centro Astron\'omico Hispano Alem\'an, C/ Jes\'us Durb\'an Rem\'on 2-2$^\circ$, 04004 Almer\'ia, Spain}
}

\author{Wolfgang Brandner}{
  address={Max-Planck-Institute for Astronomy, K\"onigstuhl 17, 69117 Heidelberg, Germany}
}

\author{Markus Janson}{
  address={University of Toronto, Department of Astronomy, St. George Street 50, M5S
3H4 Toronto, ON, Canada},
}

\author{Stefan Hippler}{
  address={Max-Planck-Institute for Astronomy, K\"onigstuhl 17, 69117 Heidelberg, Germany}
}

\author{Thomas Henning}{
  address={Max-Planck-Institute for Astronomy, K\"onigstuhl 17, 69117 Heidelberg, Germany}
}

\begin{abstract}
{\em AstraLux} is the Lucky Imaging camera for the Calar Alto 2.2-m telescope 
and the 3.5-m NTT at La Silla. It allows nearly diffraction limited imaging 
in the SDSS {\em i'} and {\em z'} bands of objects as faint as  
{\em i'}=15.5\,{\em mag} with minimum technical effort. 

One of the ongoing {\em AstraLux} observing programs is a binarity survey among 
late-type stars with spectral types K7 to M8, covering more than 1000 targets on the 
northern and southern hemisphere. The survey is designed to refine binarity 
statistics and especially the dependency of binarity fraction on spectral type. 
The choice of the SDSS {\em i'} and {\em z'} filters allows to obtain spectral 
type and mass estimates for resolved binaries. 

With an observing efficiency of typically 6 targets per hour we expect to  
complete the survey in mid-2009. Selected targets will be followed up astrometrically 
and photometrically, contributing to the calibration of  the mass-luminosity relation 
at the red end of the main sequence and at visible wavelengths.
\end{abstract}

\maketitle


\section{Introduction}
Observations of binary stars can not only constrain stellar formation theories, which have to
reproduce the observed binary frequency, but are the only way to reliably determine stellar
masses by orbit fitting techniques.
While binary statistics are well established for solar-like stars, this is not entirely true for
the low-mass and very low-mass end of the main sequence. Here large binarity surveys can
still provide substantial improvements of our knowledge of the binary fraction and its dependency
on spectral type, giving indirect evidence of a possible discontinuity in the low-mass initial
mass function \cite{Thies:2007}.

Apart from that, dynamical mass estimates of low-mass stars are the key ingredient to
the calibration of mass-spectral-type and mass-luminosity relations at the red end of
the main sequence. While stellar evolution models reproduce near-infrared magnitudes
and colours of low-mass stars quite satisfyingly nowadays, this is not the case for visible
wavelengths $<$1\,$\mu$m.

This is the motivation for the large {\em AstraLux} M~dwarf survey, designed to find and 
characterise late-spectral-type binary and multiple systems at visible wavelengths.
With more than 1000 M~dwarfs targeted by the initial survey programme, this is by
far the largest of its kind. 
The Lucky Imaging cameras {\em AstraLux} at the Calar Alto 2.2-m telescope and {\em AstraLux Sur} at ESO's NTT,
La Silla, provide the necessary observing efficiency to conduct such a survey in
relatively short time.

\section{The {\em AstraLux} Camera}
\begin{figure}
   	\includegraphics[width=13cm]{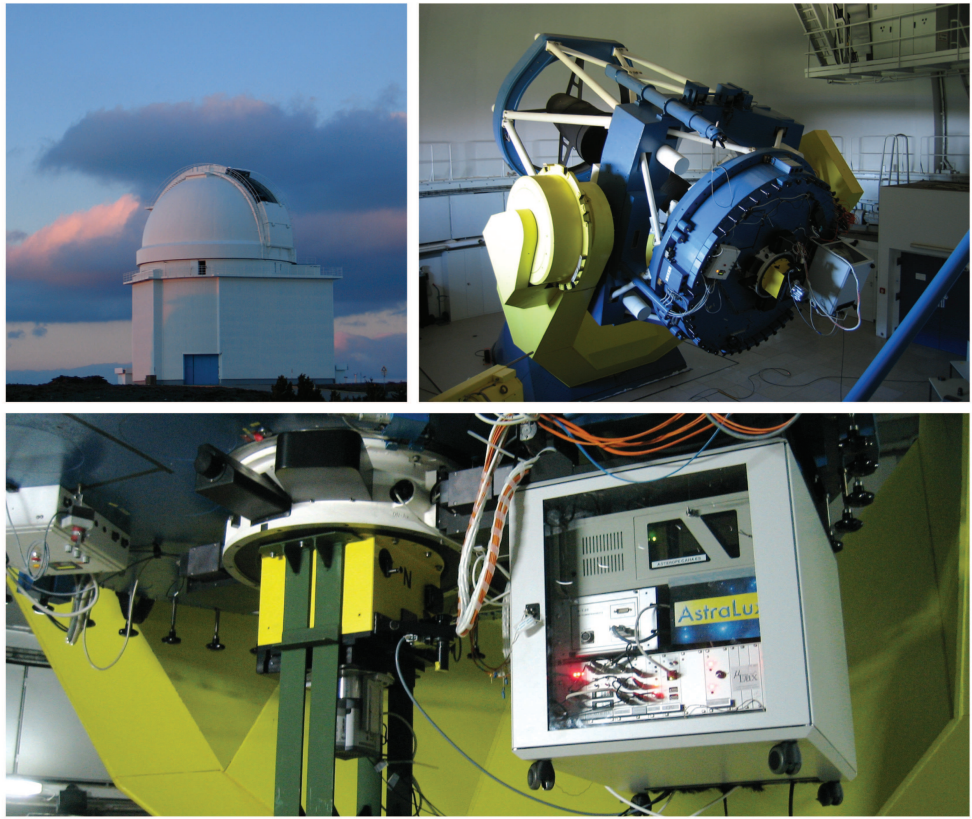}
  	\caption{ \label{fig:telescope} 
		{\em AstraLux} and the Calar Alto 2.2-m telescope. 
		{\em Top:} the 2.2-m telescope dome and the telescope
		with {\em AstraLux} attached at the Cassegrain focus. 
		{\em Bottom:} Detailed view of the instrument at the
		telescope. The yellow box contains the 8-position filterwheel, the camera is attached at the bottom.
		The grey rack to the right houses the camera control computer, a keyboard/monitor combination,
		the filterwheel electronics, and {\em MicroLux}, the GPS timing add-on.}
\end{figure}
{\em AstraLux}\footnote{\url{http://www.mpia.de/ASTRALUX/}}
 \cite{Hormuth:2008a} is a high-speed, high-sensitivity camera with an electron multiplying readout register,
allowing single-photon detection at virtually zero readout noise. Its main application is high-angular
resolution imaging employing the Lucky Imaging technique \cite{Law:2006}.
The original {\em AstraLux} camera is a common user instrument at the Calar Alto 2.2\,m telescope (see Fig.\ref{fig:telescope})
since 2007, while its sister instrument {\em AstraLux Sur} is an almost identical copy, currently used as visitor
instrument at ESO's NTT telescope at La Silla, Chile.

Both instruments allow nearly diffraction limited imaging in the SDSS {\em i'} and {\em z'} filters, i.e. at
wavelengths $<$1\,$\mu$m, with typical full width half maxima of the PSF cores below 100\,mas and
Strehl ratios of up to 20\,\%. This is achieved by recording several thousand short exposure images
(integration time $<$50\,ms) and combining only the best 5\,--10\,\% which are least affected by
blurring due to atmospheric seeing. In contrast to full-grown adaptive optics systems, this provides
high angular resolution imaging capabilities at a fraction of the technical effort and costs, and with
considerably smaller instrumental overheads.

Including target acquisition and instrument setup, typically only 10\,minutes are needed to
perform an observation of one of our survey targets in both SDSS filters. This is faster than
the overhead alone as specified for NACO, i.e. a state-of-the-art adaptive optics system.
A near real-time pipeline allows the observer to see the final high-quality results only minutes
after finishing data acquisition. 

The high frame rate of the {\em AstraLux} camera (up to several
100\,Hz using windowing and/or binning) makes the instrument also a very interesting choice
for projects needing high time resolution, e.g. transit observations, stellar occultations by solar
system objects, or pulsar timing. Accurate timing of individual exposures with an accuracy
better than 1\,$\mu$s with respect to the UTC timeframe is realised by {\em MicroLux}, a custom 
GPS based timing hardware \cite{Hormuth:2008b}.

\section{Survey Layout and Status}
\begin{figure}
   	\includegraphics[width=14cm]{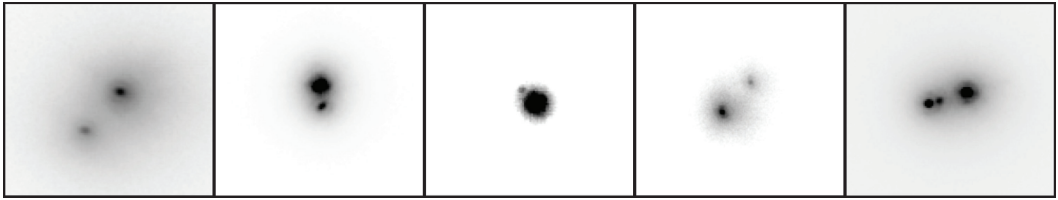}
  	\caption{ \label{fig:samples} Sample observations of M~dwarf multiples with {\em AstraLux} at the Calar Alto 2.2-m telescope. 
	The field of view is 2.5$\times$2.5\,arcsec. Effective integration times were 30\,s in the SDSS~{\em z'} filter.}
\end{figure}
Our sample comprises more than 1000 late K and M~dwarfs of spectral types K7 to M8.5.
Most of the targets were published only recently \cite{Riaz:2006,Reid:2007} and have not been
observed with high angular resolution instruments before. For most targets of spectral types earlier
than M5 spectral typing is based on spectral indices and photometric parallaxes can be assumed to
be of relatively high accuracy. Additionally, published X-ray count rates and H$\alpha$ equivalent
widths allow correlations between stellar activity and binary parameters.

\begin{figure}
   	\includegraphics[width=13.7cm]{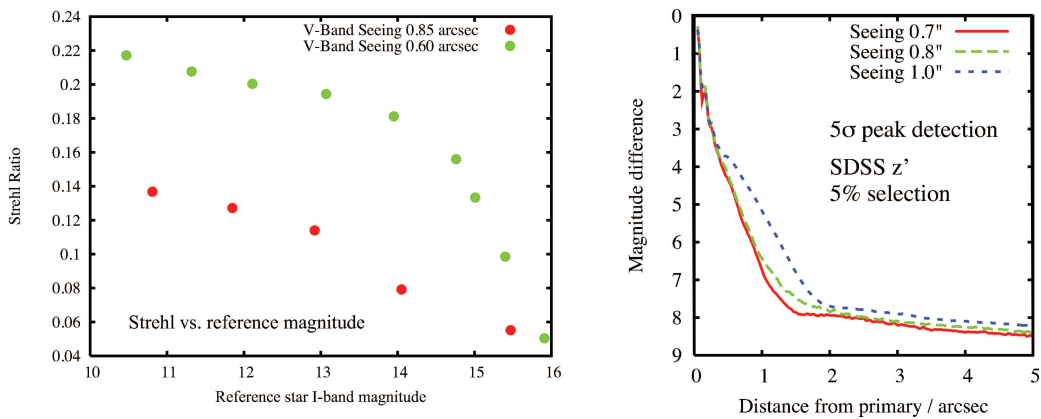}
  	\caption{ \label{fig:magdiff} 
		{\em Left: } Dependency of the final Strehl ratio on natural seeing and reference star magnitude.
		The plot is based on observations with 30\,ms single frame exposure time in the SDSS {\em z'}		
		filter and 1\% image selection rate. All Strehl ratios were measured on stars with less than 2arcsec
		separation from the reference star.	
		{\em Right:} Achievable magnitude differences for a 5$\sigma$ peak detection of a fainter companion
		to the reference star. All curves refer to a 5\% selection rate from 10000 images with 30\,ms single
		frame exposure time, equivalent to an effective integration time of 15\,s. The {\em I}-band magnitude
		of all three reference stars was $\approx$10\,mag.}
\end{figure}

All targets are initially observed in the SDSS {\em i'} and {\em z'} filters with effective
exposure times of 30\,s, allowing detection of substellar companions at distances
of up to $\approx$20\,pc. Typical imaging results of {\em AstraLux} at the Calar Alto 2.2-m
telescope are shown in Fig.~\ref{fig:samples}. Strehl ratios of more than 20\,\% can
be reached for the brighter targets of the survey under good seeing conditions, and detection
limits for close companions are generally better than e.g. compared to speckle interferometric
observations, though slightly worse than in the case of high-order adaptive optics systems (see Fig.~\ref{fig:magdiff}).

Binary parameters are extracted by PSF fitting, and obtained magnitudes and
colours are used to estimate spectral types of individual components.
All newly found binary or multiple systems will be re-observed at least once to check for
common proper motion and hence physical companionship. Systems with short period
predictions will further be observed at regular intervals to enable orbit fitting and finally dynamical
mass estimates. Additional resolved spectrophotometric observations in the near
infrared will complement the {\em AstraLux} data and allow a thorough characterisation
of these systems.

We have observed already more than 400 targets, and obtained second epoch follow-up
for a subset of these sources. We
expect to reach the benchmark of 1000 targets in mid-2009. Though full orbital solutions
will only be possible in the future, our observations will certainly allow an unprecedented refinement
of binary statistics for low-mass stars almost immediately.


\bibliographystyle{aipproc}   

\bibliography{Hormuth}

\IfFileExists{\jobname.bbl}{}
 {\typeout{}
  \typeout{******************************************}
  \typeout{** Please run "bibtex \jobname" to obtain}
  \typeout{** the bibliography and then re-run LaTeX}
  \typeout{** twice to fix the references!}
  \typeout{******************************************}
  \typeout{}
 }

\end{document}